\documentclass[12pt]{iopart}
\usepackage{float}
\usepackage{tabularx}
\usepackage{blindtext}
\usepackage{url}
\usepackage{rotating}
\usepackage{cite}
\usepackage{verbatim}
\usepackage{graphicx}

\begin{document}

\title[]{What topics of peer interactions correlate with student performance in physics courses?}

\author{L. N. Simpfendoerfer,$^1$ Meagan Sundstrom,$^{1,2}$ Matthew Dew,$^1$ and N. G. Holmes$^1$}
\address{$^1$Laboratory of Atomic and Solid State Physics, Cornell University, Ithaca, New York 14853, USA\\
$^2$Department of Physics, Drexel University, Philadelphia, Pennsylvania 19104, USA}

\ead{ngholmes@cornell.edu}

\begin{abstract}
Research suggests that interacting with more peers about physics course material is correlated with higher student performance. Some studies, however, have demonstrated that different topics of peer interactions may  correlate with their performance in different ways, or possibly not at all. In this study, we probe both the peers with whom students interact about their physics course and the particular aspects of the course material about which they interacted in six different introductory physics courses: four lecture courses and two lab courses. Drawing on methods in social network analysis, we replicate prior work demonstrating that, on average, students who interact with more peers in their physics courses have higher final course grades. Expanding on this result, we find that students discuss a wide range of aspects of course material with their peers: concepts, small-group work, assessments, lecture, and homework. We observe that in the lecture courses, interacting with peers about concepts is most strongly correlated with final course grade, with smaller correlations also arising for small-group work and homework. In the lab courses, on the other hand, small-group work is the only interaction topic that significantly correlates with final course grade. We use these findings to discuss how course structures (e.g., grading schemes and weekly course schedules) may shape student interactions and add nuance to prior work by identifying how specific types of student interactions are associated (or not) with performance.
\end{abstract}


\newpage

%
%
%
%
%

\section{Introduction}

Research has demonstrated that interacting with more peers about a science course is linked to increases in students' self-efficacy, sense of belonging, self-confidence, and academic achievement~\cite{ballen2017enhancing,sharma2005improving,bjorklund2020connections,dou2016beyond,williams2015understanding,dokuka2020academic,bruun2013talking,traxler2018networks,williams2019linking,grunspan2014}. In regard to academic achievement in particular, interacting with peers about specific course material has been shown to be central to students' learning of that material~\cite{vygotsky1978,rogoff1996,burkholder2020factors}. Peer interactions afford students the opportunity to exchange information with one another in the dynamic process of co-constructing their understanding~\cite{olitsky2007promoting,wuchty2007increasing,park2017chemical,chi2014icap}. Collaboration with others also provides opportunities for students to individually reflect on their own understanding. 

Many prior studies have performed quantitative analyses of student interactions in physics courses, largely drawing on  methods of social network analysis to both visualize and quantify peer interactions~\cite{commeford2020characterizing,commeford2021characterizing,traxler2020network,sundstrom2022intergroup,wu2022,brewe2010changing,grunspan2014,zwolak2017students,zwolak2018educational,dou2016beyond,williams2015understanding,traxler2018networks,williams2019linking,brewe2012investigating,bruun2013talking,wells2019,sundstrom2022interactions}. The subset of these studies that examine peer interactions as related to academic achievement have found that students who have more and/or stronger interaction connections to their peers tend to earn higher grades. This correlation has been observed between students' course-wide interactions and their overall course grades~\cite{williams2019linking, williams2015understanding,dokuka2020academic,grunspan2014,bruun2013talking,traxler2018networks,sundstrom2022interactions} as well as between students' lab-specific interactions and their lab grades~\cite{wei2018developing,park2017chemical}.

Most of these studies, however, examine how the \textit{number} of peer interactions in which a student engages, rather than the specific topics about which they interact, relates to student performance. One study, conducted by Bruun and colleagues~\cite{bruun2013talking}, separately analyzed networks of student interactions about problem solving and about physics concepts. They found that students who are connected to well-connected others in the problem solving network, while students who are connected to many other people in general in the physics concepts network, tend to earn higher grades. Students' interactions about different aspects of a course, therefore, may correlate with their performance in different ways, or possibly not at all. In the current study, we similarly investigate whether and how student engagement in peer interactions about various topics correlates with their performance. Different from the work by Bruun and colleagues, we determine these topics through an emergent coding scheme using students' open-ended survey responses about their peer interactions. 

We also disentangle the relationships between interaction topics and student performance in the instructional contexts of lab and lecture separately, motivated by our prior work~\cite{sundstrom2022interactions}. Given the different learning objectives, course structures, and grading schemes of these two contexts, we hypothesize that students may interact with peers about different topics in each context and that the relationship between interacting about certain topics and performing well in the course may vary between contexts. Physics labs, for example, often center around students' experimental investigations with small groups of peers, both in terms of course schedule (students often attend lab sessions for a few hours per week) and grading scheme (lab grades are often based on deliverables related to students' small-group work, such as lab reports)~\cite{hoehn2020writing}. Physics lecture courses, in contrast, place emphasis on multiple forms of course engagement: students attend both lectures and small-group problem solving sessions each week and are largely graded on exams and longer homework assignments. Thus, we both probe and analyze student interactions about these two instructional contexts separately.

This study aims to address the following research questions: 
\begin{enumerate}
    \item To what extent is the number of peers with whom a student interacts related to their final course grade in distinct lab and lecture physics courses?
    \item About what topics do students interact with their physics peers in distinct lab and lecture courses?
    \item Which, if any, peer interaction topics correlate with students' final course grades in distinct lab and lecture physics courses?
\end{enumerate}
The first question allows us to compare our data set to those in prior work by performing a similar analysis, while the second and third questions expand on the existing body of literature.

We collected survey data from six different introductory physics courses (four lecture courses and two lab courses) at Cornell University, asking students to self-report their peers with whom they interacted about the course and to describe the aspects of the course material they discussed in these interactions. We find, in most courses, that students who interact with more peers about the course material tend to earn higher final course grades, consistent with prior work. From the written explanations, we observe that the topics about which students interact with their peers closely align with the class structures: in lab, students primarily mention interacting about small-group work, while in lecture, students mostly mention interacting about homework, but they also describe interacting about specific physics concepts and lecture material. Furthermore, different topics of interactions correlate with students' final grades in lab versus lecture courses. In lab, small-group work is the only interaction topic that significantly correlates with students' final grades. In lecture, on the other hand, interacting with peers about physics concepts correlates most strongly with final course grade. Interacting about small-group work and homework also significantly correlates with final course grade in lecture, though with smaller effects than interacting about concepts. These results add nuance to prior work by identifying the specific \emph{types} of peer interactions, in addition to the \emph{number} of peer interactions, that are associated with stronger student performance.

\section{Methods}

In this section, we summarize the instructional context of our study and then describe the data collection and analysis methods.

\begin{table}[t]
\centering
\caption{\label{tab:demographics}%
Summary of survey response rates and the self-reported gender, race or ethnicity, intended major, and academic year of students in each course. Survey response rate is calculated as the percent of students enrolled in the course who completed the survey. For demographic information, the percentages are out of the number of students included in our analysis. Students are categorized as non-URM (underrepresented and minoritized) if they only self-identified as White and/or Asian or Asian American and as URM if they self-identified as at least one of the following: American Indian or Alaska Native, Black or African American, Hispanic or Latinx, and Native Hawaiian or other Pacific Islander.}
\setlength{\extrarowheight}{1pt}
\scriptsize
\vspace{0.5cm}
\begin{tabular}{lccccccc}
\hline
\textrm{}&
\multicolumn{2}{c}{Lab Course}&
\multicolumn{2}{c}{Lecture Course - Engineering}&
\multicolumn{2}{c}{Lecture Course - Physics}\\ 
\hline
\textrm{}&
\textrm{Fall}&
\textrm{Spring}&
\textrm{Fall}&
\textrm{Spring}&
\textrm{Fall}&
\textrm{Spring}\\
\hline 
Survey response rate & 95\% & 98\% & 99\% & 95\% & 90\% & 100\%\\
Students in analysis & 385 & 646 & 238 & 520 & 45 & 36\\
Gender \\
\hspace{5mm}Men & 199 (52\%) & 293 (45\%) & 112 (47\%) & 226 (43\%) & 32 (71\%) & 19 (53\%)\\
\hspace{5mm}Women & 152 (39\%) & 302 (47\%) & 116 (49\%) & 258 (50\%) & 8 (18\%) & 12 (33\%)\\
\hspace{5mm}Non-binary & 4 (1\%) & 2 (0\%) & 2 (1\%) & 2 (0\%) & 2 (4\%) & 0 (0\%)\\
\hspace{5mm}Unknown & 30 (8\%) & 49 (8\%) & 18 (8\%) & 34 (7\%) & 3 (7\%) & 5 (14\%) \\

Race or ethnicity\\
\hspace{5mm}Non-URM & 289 (75\%) & 390 (60\%) & 175 (74\%) & 307 (59\%) & 35 (78\%) & 19 (53\%)\\
\hspace{5mm}URM & 59 (15\%) & 164 (25\%) & 39 (16\%) & 144 (28\%) & 6 (13\%) & 10 (28\%)\\
\hspace{5mm}Unknown & 37 (10\%) & 92 (14\%) & 24 (10\%) & 69 (13\%) & 4 (9\%) & 7 (19\%)\\
 
Major\\
\hspace{5mm}Physics or Engineering Physics & 56 (15\%) & 51 (8\%) & 17 (7\%) & 19 (4\%) & 35 (78\%) & 27 (75\%)\\
\hspace{5mm}Engineering & 259 (67\%) & 475 (74\%) & 178 (75\%) & 409 (79\%) & 3 (7\%) & 2 (6\%)\\
\hspace{5mm}Other science & 37 (10\%) & 35 (5\%) & 20 (8\%) & 28 (5\%) & 4 (9\%) & 1 (3\%)\\
\hspace{5mm}Unknown & 33 (9\%) & 85 (13\%) & 23 (10\%) & 64 (12\%) & 3 (7\%) & 6 (17\%)\\

Academic year \\
\hspace{5mm}First & 317 (82\%) & 594 (92\%) & 190 (80\%) & 468 (90\%) & 41 (91\%) & 31 (86\%) \\
\hspace{5mm}Second & 52 (14\%) & 11 (2\%) & 37 (16\%) & 9 (2\%) & 2 (4\%) & 0 (0\%)\\
\hspace{5mm}Other or unknown & 16 (4\%) & 41 (6\%) & 11 (5\%) & 25 (5\%) & 2 (4\%) & 5 (14\%)\\
\hline
\end{tabular}
\end{table}

\subsection{Courses and participants}

The data for this study came from two offerings (fall and spring of the same academic year) of three in-person introductory physics courses -- two lecture courses and one lab course -- at Cornell University (six courses total; Table~\ref{tab:demographics}). 

Both lecture courses were calculus-based mechanics courses. One lecture course was primarily designed to serve non-physics majors (predominantly engineering students and other science majors) and the other lecture course was designed to serve physics majors. However, students were encouraged to take courses according to their preferences and academic preparation, leading to some variety in students' majors between each course (Table~\ref{tab:demographics}). The lecture course for engineers was larger, with 300-500 students per semester, and was taught in a ``flipped-classroom" format. Students in this course attended three 50 min lectures each week. Before each lecture session, students were assigned textbook readings and a short reading quiz. Each lecture section contained 150-300 students in a stadium-style lecture hall, where students engaged in active learning activities such as review of the pre-class reading material, clicker questions, and demonstrations. Students also attended two 50 min discussion sections each week. Each discussion section contained 20-25 students and was led by a graduate teaching assistant and often a supporting undergraduate teaching assistant. In discussion sections, students worked in self-selected small groups of three or four to complete ungraded practice problems. These small groups remained consistent throughout the semester. The lecture course for physics majors, in contrast, contained 30-50 students and was taught predominantly through traditional lectures in stadium-style classrooms, with a few clicker questions per lecture that students answered individually. Students in this course also attended three 50 min lectures each week. Similar to the other lecture course, students attended two 50 min discussion sections containing 20-25 students each week, where they completed ungraded practice problems in self-selected small groups (which were consistent over the course of the semester) or followed along as a graduate teaching assistant demonstrated problem solutions. Once per week, students completed a short, graded quiz during discussion. In both lecture courses, students completed independent problem sets each week for homework. Students were offered optional, course-facilitated homework sessions outside of class supported by the course instructor or graduate and undergraduate teaching assistants. Each lecture course had two midterm exams and a final exam. The grading scheme for each lecture course is shown in Table~\ref{tab:gradingschemes}. 

The lab course was offered as a distinct course (i.e., separate course code and final grade) in which the students in the two lecture courses described above typically co-enrolled. The lab course was larger than the combination of the two lecture courses (400-600 students) because students who had transfer or AP credits for the lecture course were still required to take the lab course. The lab course focused on teaching experimental skills, as in Refs.~\cite{holmes2018introductory,Smith2021,smith2020direct,holmes2015teaching}, with experiment topics spanning both mechanics and electromagnetism. Students attended one 50 min lecture each week, where each lecture section contained 200-300 students in a stadium-style lecture hall. These lectures focused on experimental and statistical analysis topics and students participated in collaborative active learning activities including small group discussion and clicker questions. Students also attended a 2 h lab session each week, which contained 20-25 students and was facilitated by a graduate teaching assistant and often a supporting undergraduate teaching assistant. Students worked in small groups of two to four to complete open-ended experimental investigations. At the end of each session, each group submitted lab notes for a group grade. Periodically throughout the semester, groups delivered oral presentations about their lab projects to the rest of their lab section. Lab groups were assigned by the graduate teaching assistant, with consideration given to student preferences indicated on an online survey at the beginning of the semester. The teaching assistants were also advised to avoid creating groups with a lone woman. Lab groups stayed consistent throughout the whole semester. 
Students took one midterm quiz and one final quiz in this course. Students also completed independent homework assignments in Jupyter Notebook focused on data analysis techniques and concepts, as well as reflection exercises about ethics, collaboration, and experimental design~\cite{fussell2022intro}. The grading scheme for the lab course is shown in Table~\ref{tab:gradingschemes}. 

Students who needed to take calculus-based introductory mechanics were advised to co-enroll in one of the two lecture courses \textit{and} the lab course. Most students co-enrolled, but it was ultimately up to the student if they would like to take both courses together. Because our data collection took place solely in the lab course, there may have been students in the lecture courses that our data collection missed. The survey response rates were all above 90\% of enrolled students, however, suggesting that the proportion of students our data collection missed was small (Table~\ref{tab:demographics}). On the other hand, between 10\% and 30\% of students in the lab course were not enrolled in one of the two lecture courses analyzed in this study. These students were likely only taking the lab course (e.g., if they had AP credit for the lecture course) or were enrolled in a different lecture course than the two lecture courses analyzed here (such as the next course in the sequence focused on electricity and magnetism).


\begin{table}[t]
\caption{\label{tab:gradingschemes} Grading schemes for each of the three courses analyzed in this study. The grading schemes were consistent between the fall and spring semesters in each course. 
}
\footnotesize
\centering
\vspace{0.5cm}
\setlength{\extrarowheight}{2pt}
\begin{tabular}{lc}
\hline 
\multicolumn{2}{c}{Lab Course} \\ \cline{1-2}
Course Component & Grade Weight (\%) \\ \hline
Lab Notes and Presentations & 54 \\
Lecture Attendance and Participation & 18 \\
Homework & 15 \\
Quizzes & 12 \\
Pre- and Post-survey Completion & 1 \\
\hline
\multicolumn{2}{c}{Lecture Course - Engineering} \\ \cline{1-2}
Course Component & Grade Weight (\%) \\ \hline
Final Exam & 25 \\
Midterm 1 & 20 \\
Midterm 2 & 20 \\
Discussion & 20 \\
Reading Quizzes & 5 \\
Lecture & 5 \\
Homework & 5 \\
\hline
\multicolumn{2}{c}{Lecture Course - Physics} \\ \cline{1-2}
Course Component & Grade Weight (\%) \\ \hline
Homework and Quizzes & 30 \\
Final Exam & 30 \\
Midterm 1 & 20 \\
Midterm 2 & 20 \\
\hline
\end{tabular}
\end{table}

\subsection{Data collection}

We administered an online survey via Qualtrics as part of a homework assignment in the lab course. The survey was given in the middle of the 15-week semester, after students had completed at least one assessment in both the lab and lecture courses. 
The survey asked students to list peers in each of their physics courses (lab and lecture) with whom they recently had a meaningful interaction using a survey prompt adapted from prior work~\cite{zwolak2018educational,traxler2020network,dou2019practitioner,commeford2021characterizing,sundstrom2022interactions}. For each peer that they listed, students were also asked to explain what aspects of the course material they discussed with that peer. The survey prompts were as follows:
\begin{quote}
    \textit{Please list any students in your physics [lab or lecture] class that you had a meaningful interaction* with about course material this week. What aspects of the course material did you discuss with this person?}\\
    \\
    \textit{*A meaningful interaction may mean in class, out of class, in office hours, virtually, through remote chat or discussions boards, or any other form of communication, even if you were not the main person speaking or contributing.}
\end{quote}
Each question was in an open-response format, with separate text boxes to enter each name and each explanation about an interaction. Students could nominate up to 15 peers for each question, but no student nominated close to 15 others. As mentioned in prior work, ``students self-identified what counted as a meaningful interaction"~\cite[p. 6]{commeford2020characterizing}. We asked students about who they interacted with ``this week" to capture interactions that students were regularly having with their peers throughout the course while reducing the possibility of recall bias (e.g., if we asked them to recall all peers with whom they have interacted throughout the semester). This phrasing may have captured a few one-off interactions that only occurred the week of the survey, however such interactions likely represent a small fraction of the reported interactions.

In our data cleaning, responses containing misspelled names, nicknames, or only a first or last name were manually compared to the course roster by the first and second authors. If only a first or last name was listed that was not unique within the course roster, that particular interaction was dropped from the analysis. We were able to match at least 90\% of all nominations made in the survey to the roster. Additionally, response rates for each course in the analysis were at least 90\% (Table~\ref{tab:demographics}). Applying methods of social network analysis to this data set, therefore, is reliable because we have less than 30\% missing data~\cite{smith2013structural}.

The survey also asked students to self-disclose their demographic information, including gender, race or ethnicity, academic major, and year (Table~\ref{tab:demographics}). Both the lecture course for engineers and the lab course contained approximately even proportions of men and women, while a majority of the students in the lecture course for physics majors were men. Additionally, the composition of underrepresented and minoritized (URM) students doubled from approximately 15\% to 30\% between the fall and spring offerings of each of the three analyzed courses. The lab course contained a majority of engineering students. In the lecture courses, the composition of academic majors generally followed the expectation (i.e., majority of engineering majors in the lecture course for engineers and majority of physics majors in the lecture course for physics majors), with slight variation between semesters. All six courses also contained at least 80\% first-year students.

\subsection{Data analysis}

We conducted three stages of analysis: determining the structure of the peer interaction networks, measuring the relationship between students' position in the interaction networks and their course grade, and identifying which kinds of interactions are correlated (or not) with students' course grades.

\subsubsection{Network structure}

We first used methods of social network analysis~\cite{grunspan2014,dou2019practitioner,brewe2018guide} to understand the broad structural features of our six interaction networks (one for each of two offerings of the three courses). We converted the self-reported peer interactions into directed networks. Each student was considered a \textit{node} in the network and each reported interaction from the survey was considered an \textit{edge}. Edges pointed from the nominating student to the student with whom they reported an interaction. A one-way edge indicated that one student reported having a meaningful interaction with another student, while a two-way edge indicated that two students reported having a meaningful interaction with each other. Interactions conceptually indicate mutual communication and involvement, thus one could interpret all reported interactions as inherently two-way edges. However, one-way interactions may indicate two forms of survey bias. One possibility is recall bias, where one student does not remember the interaction or the name of the other student. The other possibility is an over-reporting of interactions, where the nominator listed many cursory interactions and the nominated student did not consider the interaction meaningful. These biases respectively produce an under-representation or over-representation of meaningful interactions in the course. We treated the networks as directed in our analysis, therefore, such that mutually reported edges are weighted more than one-way edges, but all reported edges are still considered. This treatment helped provide a middle-ground between both possible biases.


For each network, we measured four different network statistics to describe the overall structure:
\begin{enumerate}
    \itemsep0em 
    \item \textit{Density}: the number of edges in the observed network as a proportion of all possible edges that could exist in the network
    \item \textit{Transitivity}: the tendency of nodes in the network to cluster together, measured as the proportion of two-paths (two edges connecting three nodes) that are closed by a third edge to form a triangle
    \item \textit{Number of clusters}: the number of groups of nodes that are connected to each other but not to any other node in the network
    \item \textit{Giant component}: the number of nodes contained in the largest cluster of the network
    \item \textit{Number of isolates}: the number of nodes in the network that are not connected to any other nodes (i.e., nodes with zero adjacent edges)
\end{enumerate}

We determined the standard errors of the density and transitivity values via bootstrapping with the \textit{snowboot} package in R~\cite{chen2019snowboot} to get a sense of uncertainty in the statistics. Each observed network was re-sampled in 10,000 bootstrap trials. We calculated a given network statistic for each sampled network and found the standard error of the statistic across the distribution of all sampled networks.

\subsubsection{Relationship between interaction networks and grades}

Next, we used exponential random graph models (ERGMs) to understand the relationship between students' number of peer interactions and their final grades, controlling for other measurable variables in the networks. ERGMs assume that networks form from a series of social processes and that this formation is often related to the qualities of the members of the network. Therefore, the model considers ways that the nodes of a network might self-organize on a structural level and how attributes of those nodes (e.g., grades) are related to the way they organize. 

The model assumes that an observed network is one of a large exponential distribution of potential networks that could form from the given set of nodes. The ERGM models this distribution of possible network structures and determines if patterns of organization (i.e., students with higher grades having more central positions in the network) are significantly more present in the observed network than would occur by random chance~\cite{anderson1999,robins2007intro}. The goal is to use $k$ predictor variables or network statistics, $g_k(y)$, and their corresponding coefficients $\theta_k$ to predict the structure of the random (observed) network $Y$. The model takes the form:
\begin{equation*}
    P_\theta[Y = y] = \frac{1}{\psi}\exp\left(\sum_{k} \theta_k g_k(y)\right)
\end{equation*}
where $y$ is a realization of the random network $Y$ and $\psi = \sum_y \exp\left(\sum_{k}\theta_k g_k(y)\right)$ is a normalization constant that ensures that the probability sums to one. Given an observed network $y$, the coefficients of the model are estimated using Maximum Likelihood Estimation (MLE). Due to the dependence between the network edges, the MLE is commonly approximated with Markov Chain Monte Carlo (MCMC) techniques~\cite{hunter2008}. The coefficients $\theta_k$ represent log-odds of tie formation and can be interpreted as a weighting of the importance of each modeled configuration for the realized network, where positive coefficients show that the configuration is observed more frequently than by chance after accounting for all other configurations that are modeled, and vice versa for negative coefficients. 

We chose a set of predictor variables that incorporated both structural variables and nodal variables, similar to our prior work~\cite{sundstrom2022interactions}. 
Our final model included the following predictor variables:



\begin{enumerate}
    \itemsep0em 
    \item \textit{Edges}: main intercept term measuring the number of observed edges
    \item \textit{Reciprocity}: measure of reciprocal or two-way edges (e.g., student A reports an interaction with student B and student B reports an interaction with student A)
    \item \textit{Geometrically-weighted out-degree (GWOD)}; decay parameter = 0.7: measure of the distribution of outgoing edges of each node 
    \item \textit{Homophily on lab section}: measure of edges occurring between students in the same lab section
    \item \textit{Homophily on discussion section}: measure of edges occurring between students in the same discussion section
    \item \textit{Homophily on lab group}: measure of edges occurring between students in the same lab group
    \item \textit{Homophily on gender}: measure of edges occurring between students of the same gender
    \item \textit{Main effect of gender on degree (woman)}: measure comparing women's total number of adjacent edges to men's total number of adjacent edges
    \item\textit{Homophily on race or ethnicity}: measure of edges occurring between students of the same URM status
    \item \textit{Main effect of race or ethnicity on degree (URM)}: measure comparing URM students' total number of adjacent edges to non-URM students' total number of adjacent edges
    \item \textit{Main effect of final course grade on degree}: measure of correlation between students' total number of adjacent edges and their final course grade 
\end{enumerate}
The first three predictor variables measured structural features of each network, while the remaining eight predictor variables measured the relationship of node-level attributes to the formation edges of the network. In this study, we focused on the \textit{main effect of final grade on degree} variable to investigate the relationship between students' position in the interaction networks and their performance in the course. We included the other predictor variables to control for other aspects of students' identities and participation that likely affect network formation. Removing these terms from the model may have led to different results for the relationship
between grade and network degree through omitted variable bias~\cite{walsh2021omittedbias}. 

We note that this list of variables differs slightly from that used in our previous work~\cite{sundstrom2022interactions}. First, we added a term to measure the tendency for students to nominate peers in their same lab group, beyond just those in their same lab or discussion section. We also found that models using the term for geometrically-weighted edgewise shared partners (GWESP) did not converge for our observed networks. Thus, we replaced this term with the geometrically-weighted outdegree (GWOD) term, a term similar to GWESP that aids in model convergence and preventing model degeneracy~\cite{goodreau2007}. The GWOD term accounts for the outdegree (the number of adjacent outgoing edges to a node, indicating number of nominations reported on the survey) distribution for all nodes in the network, with more weight placed on nodes with lower outdegrees (lower numbers of nominations reported on the survey) because such distributions are often highly skewed~\cite{robins2007recentdevs,hunter2007}. Including this term allowed for model convergence and improvements to the goodness-of-fit diagnostics (see Fig.~\ref{fig:gofplots} in the Appendix) because our observed networks had a large proportion of students who did not nominate others. 

We also note that the \textit{main effect of final course grade on degree} variable cannot handle missing data, therefore nodes and associated edges that are missing grade data (i.e., students that may have dropped or withdrawn from the course after completing the network survey) were not included in the ERGM analysis. In all cases, at least 90\% of enrolled students were retained in the ERGM analysis.

\subsubsection{Types of interactions and their relationship with grades}

To further examine the correlation between student interactions and their final course grades, we analyzed students' responses to the question: ``What aspects of the course material did you discuss with this person?" We performed a thematic coding analysis to identify the main aspects of the courses being discussed among students. The first and third authors initially read all of the student responses across the six analyzed courses to get a sense of the data as a whole~\cite{Tesch1990}. These authors then identified common themes in the explanations and defined a preliminary codebook. The same authors then iteratively coded a subset of responses independently, met to discuss coding disagreements, and modified code definitions~\cite{Campbell2013}. Modifications to the codebook were also discussed with the full project team between iterations. 

Once the coding scheme was finalized, the two authors coded a random sample of 100 of the 1,982 total reported interactions (1,177 in lab and 805 in lecture) across all six courses to determine interrater reliability. We stratified the random sample by instructional context (50 in lab and 50 in lecture) because the courses were structured differently and had different learning objectives. We calculated Fuzzy Kappa~\cite{kirilenko2016inter} to determine interrater reliability between the two coders because multiple codes could be applied to each response. Fuzzy Kappa was 0.93, exceeding the reliability threshold of 0.80~\cite{kirilenko2016inter}. After establishing reliability, the first author coded the remaining explanations.

We then created histograms of the code frequencies to determine which aspects of the lecture and lab courses students interacted with each other about the most. When determining the frequencies of each code, each reported interaction was counted separately. That is, if two students mutually reported an interaction with one another, the codes from each of their reported interactions were counted. We also combined data from the lecture course for physics majors and the lecture course for engineers because there were not large differences in the code distributions in each of these courses. Within each instructional context, lab and lecture, we aggregated the data from the fall and spring offerings because there were not substantial differences in the code distributions when considering each offering separately. This larger data set helped to reduce possible noise in our statistical analysis. 

We employed linear mixed models, or hierarchical linear models~\cite{vandusen2019hierarchical}, to understand how the topics of students' interactions, identified by our coding scheme, related to their course performance. We ran a linear mixed model for each context, lab and lecture, because the relationships between interaction topics and final grades may be different in each context. We included a random effect in each model to account for variations (e.g., instructor, instructional style, and students' prior preparation) in the different lecture courses students were taking: fall offering of the lecture course for engineers, fall offering of the lecture course for physics majors, fall offering of any other lecture course, spring offering of the lecture course for engineers, spring offering of the lecture course for physics majors, and spring offering of any other lecture course. The ``any other lecture course" categories were only used in the lab model, where there were students who were not enrolled in one of the two lecture courses analyzed in this study. We calculated the intraclass correlation coefficient (ICC) for each model to verify the inclusion of this random effect~\cite{vandusen2019hierarchical}. ICC quantifies the fraction of the total variance in the student-level data that can be attributed to variance between each offering of the lecture courses. Typically, a random effect should be included in the model if the ICC for that effect is at least 0.05. The ICC for course was 0.02 and 0.11 in the lab and lecture models, respectively. Thus, the ICC for the lecture model surpasses the common threshold, while the ICC for the lab model does not. We opted to use the same linear mixed model for both the lab and lecture contexts for consistency. We also checked that using a single-level linear regression (i.e., without the random effect) for the lab context produced the same overall results as the linear mixed model. 



In the models, students' final grades were the dependent variable and whether or not they interacted with at least one peer about each interaction code comprised the binary predictor variables. Students' final course grades were converted from letter grades to grade point average (GPA) points. We found that this produced enough discrete values to be approximated as a continuous variable. We also checked that using an ordinal logistic regression produced the same overall results, but report the results of the linear regression because they are more interpretable. 

For the predictor variables, we considered a student as interacting about a given code if they had at least one interaction (either an incoming or outgoing edge) that received that code. We decided not to consider the number of times each student interacted about each code because the distributions of codes per student were highly skewed, with many students having zero interactions and very few students having at least two interactions about each code. Grouping together students who had at least one interaction about each code, therefore, allowed for more comparable sample sizes of students with zero and at least one interaction. Additionally, considering the number of times each student interacted about each code would simultaneously measure the effect of the number of interactions the student had and the topic of interaction. Considering only whether the student had at least one interaction with each code more explicitly addressed our third research question. That is, this treatment of the predictor variables directly determined whether or not students who interact with peers about a given topic, regardless of the number of these peers, tend to receive a higher course grade than students who do not interact with any peers about that topic.

Similar to our ERGMs, the linear mixed models did not include students with no final course grade (less than 10\% of enrolled students). Interactions reported by a student who was dropped from the analysis, however, still counted for the student who remained in the analysis.

\begin{figure*}
    \centering
    \includegraphics[width=5.5in]{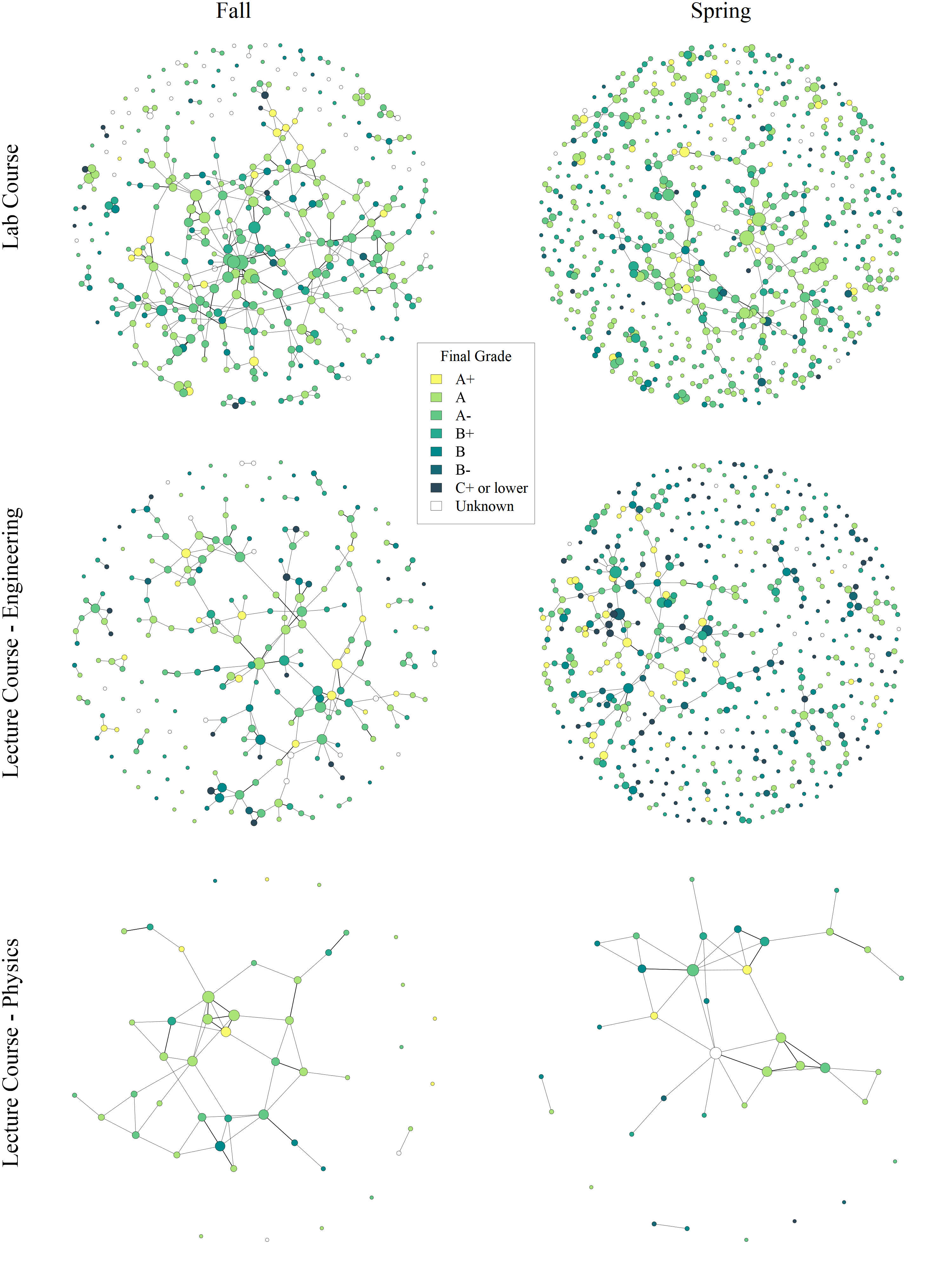}
    \caption{Diagrams of interaction networks for all six courses. Nodes are colored by final course grade and sized proportional to total degree (number of edges connected to each node). Thick edges represent reciprocal edges (students A and B both reported interacting with one another) and thin edges represent one-way edges (student A reported interacting with student B, but student B did not report interacting with student A).}
    \label{fig:sociograms}
\end{figure*}

\begin{table}[t]
\caption{\label{tab:networkstats}%
Summary of network-level statistics for the observed interaction networks. Standard errors of the last digit of density and transitivity are shown in parentheses. The percentages for size of giant component and isolates are calculated as a fraction of all nodes in a given network.}
\vspace{0.5cm}
\scriptsize
\centering
\setlength{\extrarowheight}{1pt}
\begin{tabular}{lccccccc}
\hline
\textrm{}&
\multicolumn{2}{c}{Lab Course}&
\multicolumn{2}{c}{Lecture Course - Engineering}&
\multicolumn{2}{c}{Lecture Course - Physics}\\ \hline
\textrm{}&
\textrm{Fall}&
\textrm{Spring}&
\textrm{Fall}&
\textrm{Spring}&
\textrm{Fall}&
\textrm{Spring}\\
\hline 
Density & 0.003(3) & 0.002(1) & 0.004(5) & 0.001(1) & 0.03(7) & 0.04(1)\\
Transitivity & 0.3(2) & 0.3(2) & 0.2(2) & 0.2(2) & 0.3(7) & 0.3(8) \\
Number of clusters & 21 & 73 & 18 & 51 & 2 & 3 \\
Giant component & 261 (67\%) & 238 (36\%) & 146 (61\%) & 179 (34\%) & 32 (70\%) & 27 (75\%) \\
Number of isolates & 65 (17\%) & 120 (18\%) & 46 (19\%) & 158 (30\%) & 12 (26\%) & 5 (14\%) \\
\hline
\end{tabular}
\end{table}

\section{Results}

In this section, we present the results for each stage of analysis.

\subsection{Network structure}

The network diagrams and network-level statistics for all six interaction networks are shown in Fig.~\ref{fig:sociograms} and Table~\ref{tab:networkstats}, respectively. In Fig.~\ref{fig:sociograms}, each student is represented as a node and the nodes are colored by students' final course grades, with darker blue indicating lower grades and lighter green and yellow indicating higher grades. Nodes are also sized by total degree (sum of incoming and outgoing edges), with larger nodes having more connections in the network than smaller nodes. Each of the connections (edges) between nodes represent a reported interaction, with thin lines representing one-way edges (only one student reported the interaction) and thick lines representing two-way edges (both students reported the interaction). 

We observe that most of the networks are quite interconnected, containing many edges. Network densities cannot be directly compared across networks of vastly different sizes because this measure does not scale linearly with the number of nodes in a network, however we see that the densities for both offerings of the lab course and the lecture course for engineers (courses with relatively similar class sizes) are comparable in magnitude. The densities of interaction networks in the lecture course for physics majors are larger in magnitude because there are fewer nodes and thus fewer possible edges. 

The networks also contain relatively large giant components (the largest interconnected cluster of students in a given network). Specifically, in four out of six courses, more than 60\% of all students in the network are connected within this giant component. We suspect that the smaller proportion of students in the giant component in the spring offerings of the lab course and the lecture course for engineers is due to the class sizes increasing by 1.5 times in the lab course and 2.2 times in the lecture course compared to the fall offerings. Indeed, the raw number of students in the giant components of these two networks is comparable to the number of students in the giant components of the fall offerings. Similarly, students in the spring offerings of the lab course and the lecture course for engineers form more individual clusters (more than 50) than students in the other four courses (21 or fewer). These observations are likely due to students having a fixed capacity for knowing and interacting with peers regardless of the size of the class in which they are enrolled. 

\begin{figure}[t]
    \centering
    \includegraphics[width=3in,trim={0 0 7in 0}]{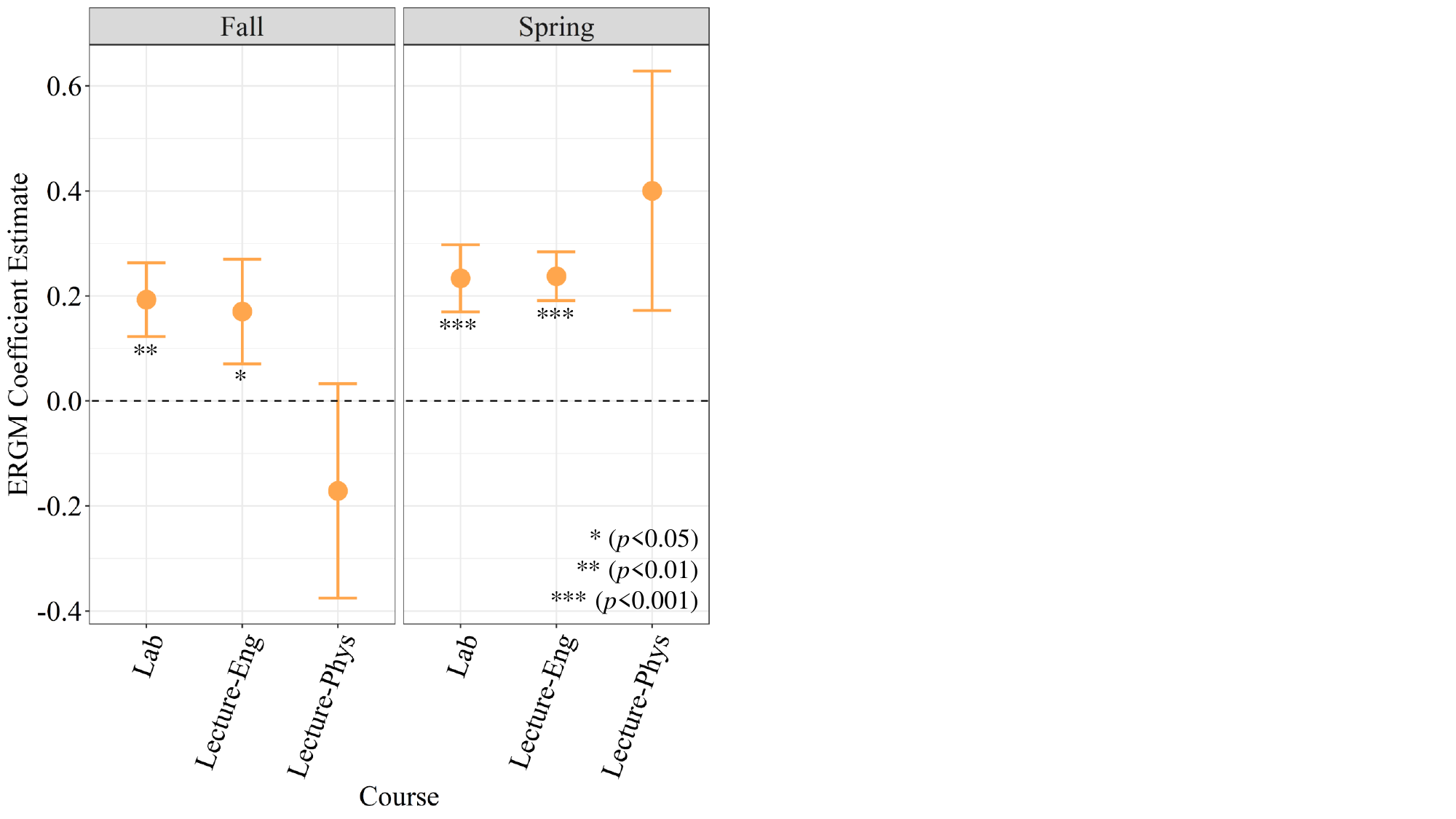}
    \caption{Plot of ERGM coefficients for the \textit{main effect of final course grade on degree} variable for each observed network. A more positive (negative) coefficient estimate indicates that students with higher final course grades have more (fewer) total connections in the network than students with lower final course grades. Error bars indicate the standard error for each estimate and asterisks indicate statistical significance.}
    \label{fig:gradeplot}
\end{figure}

\begin{table}
\caption{\label{tab:codingscheme}%
Definitions and examples of the coding scheme characterizing students' explanations of what they talk about with other students. The same codes were applied in all courses.}
\footnotesize
\centering
\vspace{0.5cm}
\setlength{\extrarowheight}{5pt}
\begin{tabular}{p{0.12\linewidth} p{0.4\linewidth} p{0.4\linewidth}}
\hline
\textrm{Code}&
\textrm{Definition}&
\textrm{Example Responses}\\
\hline
Concepts & Student describes interacting about a particular physics or statistical concept without mentioning a specific assignment or activity. & “We discussed the equations for uncertainties in mean”, “Class content”, “We discussed physics concepts” \\
Small-group work & Student describes interacting about or in an aspect of the course that occurs in a small group setting, including lab and discussion sections. & “The handouts in the discussion sessions”, “Worked together in lab group” \\
Assessments & Student describes interacting about an assessment (e.g., exams and quizzes), either studying for the assessment or reflecting on it together. & “We talked about how we felt about the [exam]”, “Studying for the quiz together”\\
Lecture & Student describes interacting during lecture, such as during clicker questions or other active learning components. & “We helped each other with the iClicker questions during lecture”\\
Homework & Student describes interacting about the homework assignments. This also includes references to study halls or office hours, where students work on the homework. & “Attended [the] study hall and discussed strategies for the homework”, “We live together so we make sure to discuss the reading chapters.” \\
Other & Response is vague, captures ideas that are too infrequent to warrant a separate code, or is entirely blank. &  ``General well-being" 
\\
\hline 
\end{tabular}
\end{table}

The transitivity values are also similar across all six networks, indicating that students in all of the analyzed courses have similar tendencies to form small groups of peers with whom they interact. The proportions of nodes that are isolates are also similar across all six networks, with 14\% to 30\% of students in each course having zero adjacent edges. Isolated students did not report any interactions on the survey and no other students in the class reported interacting with them.

\subsection{Relationship between interaction networks and grades}

Visually, we observe in Fig.~\ref{fig:sociograms} that many of the large, well-connected nodes in these networks tend to have lighter colors, indicating higher grades. There are, however, some exceptions to this general trend. For example, the spring offering of the lecture course for engineers contains many large dark blue nodes, indicating lower grades, that are connected to many other nodes. 

The ERGMs allow us to quantitatively measure this relationship between students' position in the interaction network and their final grade. Controlling for other structural features of the network, we find that there is a significant, positive correlation between students' degree in the interaction network and their final course grade in both offerings of the lab course and both offerings of the lecture course for engineers (Fig.~\ref{fig:gradeplot} and Table~\ref{tab:coefficients} in the Appendix). 
In both offerings of the lecture course for physics majors, however, we see no significant correlation between students' grades and their interaction network degree.



\subsection{Types of interactions and their relationship with grades}

\begin{figure*}[t]
    \centering
    \includegraphics[width=3in,trim={.6in 0 7in 0}]{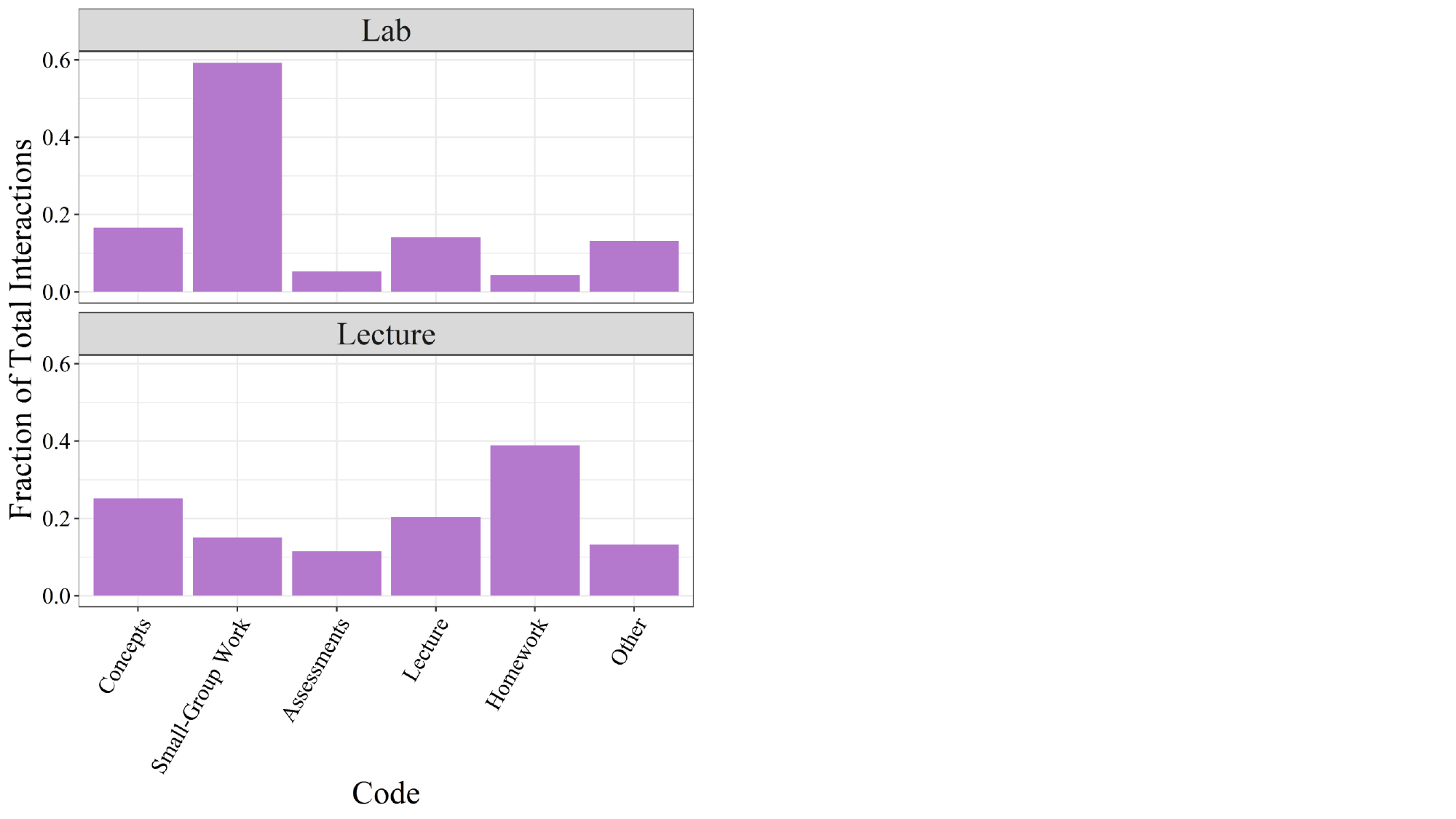}
        \caption{Histograms showing the frequencies of each interaction topic in each instructional context. The bars within each context may add up to more than one because each explanation could receive more than one code.}
        \label{fig:codedistns}
\end{figure*}

\begin{figure*}[t]
    \centering
    \includegraphics[width=3.3in,trim={0.3in 0 7in 0}]{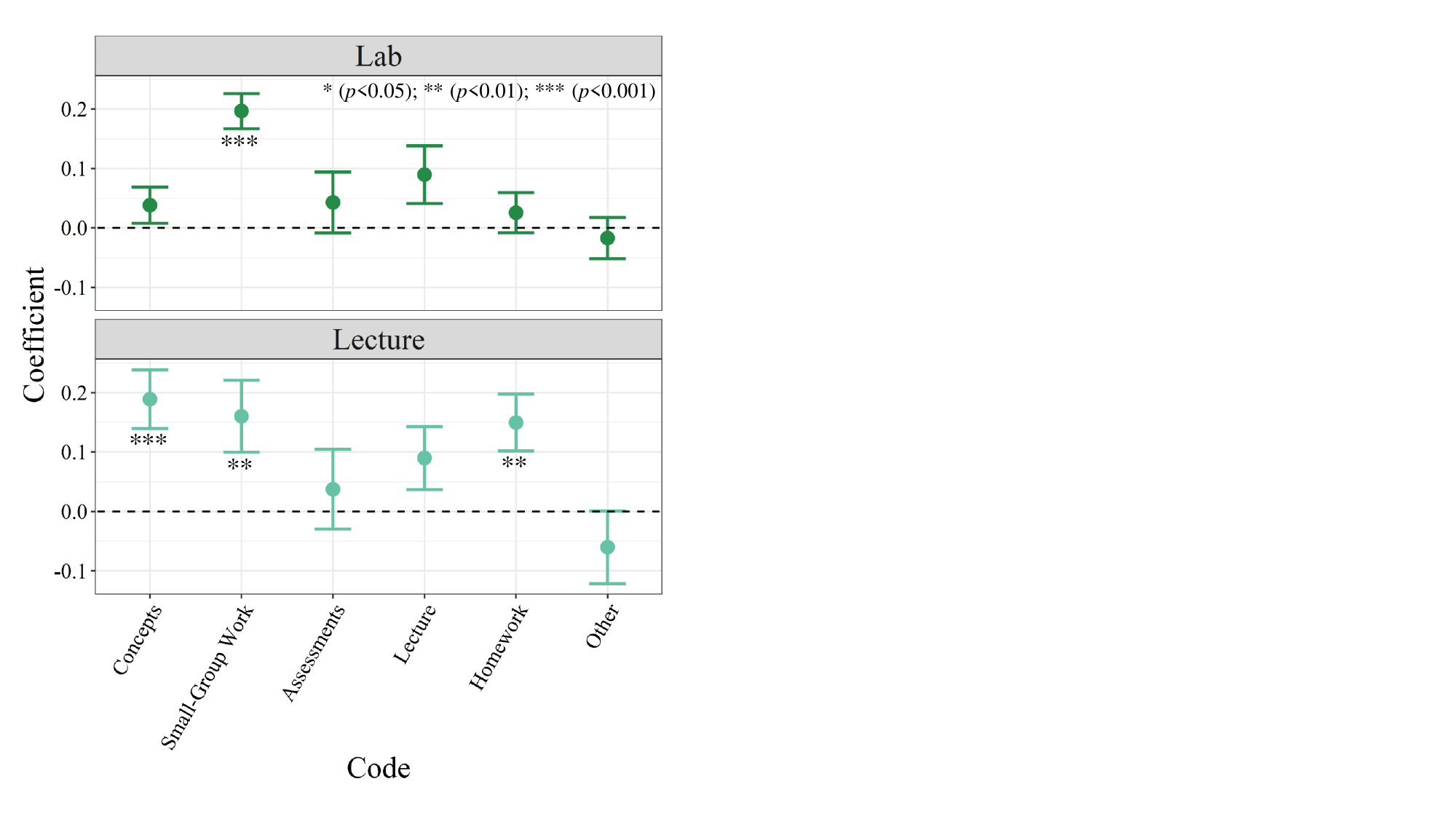}
        \caption{Linear mixed model results. Each coefficient compares the final course grades (on a GPA scale) of students that interact with at least one peer about a given topic to the grades of students who do not interact with any peers about that topic.}
        \label{fig:linearreg}
\end{figure*}

Our coding scheme captures common topics that students report interacting about with their peers (Table \ref{tab:codingscheme}). The coding scheme closely aligns with the class structure, assignments, and grading scheme (Table~\ref{tab:gradingschemes}) of each course (i.e., \textit{small-group work}, \textit{assessments}, \textit{lecture}, and \textit{homework}), with the exception of \textit{concepts}, which indicates interactions about specific physics content students are learning. 


In the lab courses (top panel of Fig.~\ref{fig:codedistns}), \textit{small-group work} is the most common interaction topic, with students mentioning this topic in about 60\% of the explanations. Students also mention \textit{concepts}, \textit{lecture}, and \textit{other} topics in 10\%-20\% of the reported interactions. In the lecture courses (bottom panel of Fig.~\ref{fig:codedistns}), interactions about \textit{homework} occur the most often (about 40\%), while the remaining five codes are similarly frequent (between 10\% and 25\%).


Linear mixed models indicate which of these interaction topics are correlated with students' final grades in each course. In the lab course, interacting about \textit{small-group work} correlates most strongly with students' final grades (top panel of Fig.~\ref{fig:linearreg} and Table~\ref{tab:linearregresults} in the Appendix). Interacting with at least one person about \textit{small-group work} corresponds to an increase in final course grade by 0.2 GPA points as compared with students who did not interact with any peers about \textit{small-group work}. Interacting with at least one peer about the lab \textit{lecture} also positively correlates with students' final grades (corresponding to an increase in final course grade by about 0.1 GPA points), though the effect is not statistically distinguishable from zero. 

In the lecture courses, we see a wider range of interaction topics that significantly correlate with students' final course grades (bottom panel of Fig.~\ref{fig:linearreg} and Table~\ref{tab:linearregresults} in the Appendix). Interacting with at least one peer about \textit{concepts} correlates most strongly with students' final grades in the lecture courses, corresponding to an increase in final course grade by about 0.2 GPA points as compared with students who did not interact with any peers about \textit{concepts}. Additionally, interacting with at least one peer about \textit{small-group work} or \textit{homework} positively correlates with final course grade, each corresponding to an increase in final course grade by about 0.15 GPA points. Similar to the lab courses, interacting with peers about \textit{lecture} positively correlates with final course grade (corresponding to an increase in final course grade by about 0.1 GPA points) in the lecture courses, but this effect is not statistically distinguishable from zero.

Interestingly, the extent to which interaction topics correlate with final course grade (Fig.~\ref{fig:linearreg}) mirrors the frequencies of the topics (Fig.~\ref{fig:codedistns}) in the lab course, with more frequently mentioned topics generally having a higher correlation with final grade in this course. In the lecture courses, on the other hand, the interaction topic frequencies do not closely mirror their relationship with student grades. While there is a breadth of interaction topics that students frequently discuss and that significantly correlate with grades in the lecture courses, there are some discrepancies to this pattern. \textit{Homework}, for example, is by far the most discussed topic, but is only moderately correlated with grades.


\section{Discussion}

In this study, we identified the specific course topics about which students interact with their peers and determined whether and how engagement in interactions about these topics is correlated with students' final course grades. In the remainder of this section, we synthesize the findings for our three research questions within each instructional context, lab and lecture, separately and then discuss the limitations of our study.


\subsection{Relationship between peer interactions and student performance in lab}

Regarding our first research question, we replicate prior work finding that engagement in more peer interactions within a lab course is positively correlated with students' lab grades~\cite{wei2018developing, park2017chemical}. We did not find such a correlation in our previous work investigating remote physics labs, possibly because the lab material was part of the larger lecture course, rather than a distinct course, and only accounted for between 10\% and 20\% of students' final course grades~\cite{sundstrom2022interactions}. The current result is still somewhat surprising because we expected, due to the collaborative nature of the lab assignments, that all students would interact with each other during in-person lab sessions. A small range in students' number of peer interactions would reduce the possibility of finding a statistically significant correlation between interactions and grades. Instead, however, our results point to the variability in the extent to which students engage in and perceive meaningful small group interactions during lab. For example, there are students who may show up to a different section than their own on any given week (e.g., due to illness) and work with an entirely new group of peers with whom they are not familiar. There may also be situations where group dynamics lead to students being excluded from the conversation. Still other students may choose to disengage from the activity entirely and opt to do other work on their phones or laptops.

Regarding our second and third research questions, we build on prior work by also examining the topics about which students interact related to lab instruction and how those interaction topics relate to student performance. Our analysis unveiled that the majority of peer interactions in the lab course were related to the \textit{small-group work} that takes place during lab sessions, as one might expect. To a much lesser extent, students also talked to their lab peers about physics \textit{concepts}, lab \textit{lecture}, \textit{assessments}, \textit{homework}, and \textit{other} topics. These topic frequencies mirrored the strength of correlations between students interacting about a given topic and their final course grade: \textit{small-group work} was the only topic that significantly correlated with student performance in the lab course. 

These results are likely due to features of the lab course structure such as the time allotted to each course component, the grading scheme, and the nature of the learning activities. In this lab course, in-class time was primarily allotted to small-group work (2 h per week), and lab notes and presentations completed during this small-group work comprised 54\% of students' final course grades (see Table~\ref{tab:gradingschemes}). The open-ended nature of the experimental investigations performed during lab sessions also necessitated peer interactions to, for example, make experimental decisions and write up and submit the lab notes for a group grade, moreso than in a traditional physics lab~\cite{sundstrom2022intergroup}. The large amount of class time and large fraction of the final course grade dedicated to small-group work, and the collaborative nature of the small-group work, therefore, likely explain why this topic was both the most common and the most strongly correlated with student performance in the lab course.

In contrast, students only attended lab lectures for 50 min per week and lecture attendance and participation only accounted for 18\% of students' final course grades. While the lectures made use of active learning strategies such as clicker questions, the collaborative nature of lecture activities likely did not outweigh the relatively low time and grade weight allotted to this course component. Likewise, homework and assessments (quizzes) did not take up a lot of time relative to other coursework, made up a small fraction of the course grade (15\% and 12\%, respectively), and were mostly completed individually.

Though not directly reflected in the course structure or grading scheme, students did not report interacting about concepts very frequently in the lab course nor did interacting about concepts correlate with student performance. With regard to physics concepts (e.g., angular momentum), this finding is consistent with the learning goals of the lab course, which focused on experimental skills rather than content reinforcement~\cite{holmes2018introductory,Smith2021,smith2020direct,holmes2015teaching}. The \textit{concepts} code, however, also captured concepts specific to the lab course, such as data analysis concepts, that students applied in their small-group work, homework, and assessments.  We suspect that students' interactions about such concepts may have been conflated in their reports of these other interaction topics (e.g., students may have interacted about concepts during the small-group work to which they referred in their written explanation). Alternatively, the lack of correlation may again reflect the overall grading scheme, where assessment of these concepts (through homework and quizzes) made up a relatively small proportion of students' grades.


\subsection{Relationship between peer interactions and student performance in lecture}

Regarding the first research question, we again replicate previous findings that students' number of peer interactions is positively correlated with their course grades in the lecture course for engineers~\cite{williams2019linking,williams2015understanding,dokuka2020academic,grunspan2014,bruun2013talking,traxler2018networks,sundstrom2022interactions}. Interestingly, however, this effect was not statistically distinguishable from zero in the lecture course for physics majors. An instinctive explanation is that this result is attributable to small sample sizes: the lecture course for physics majors contained 45 and 36 students in the fall and spring, respectively. These samples were likely sufficiently large, however, because we found other statistically significant relationships in the ERGMs for these networks. For example, we found significant gender effects in the fall lecture course for physics majors despite the small fraction of women in that course (\textit{main effect of gender on degree} variable in Table~\ref{tab:coefficients} in the Appendix).

Instead, our findings for this course could be due to range restriction, such as due to low variability in either students' number of peer interactions or students' final course grade, which would reduce the possibility of finding a significant correlation between the two variables. The range of students' network degree in the lecture courses for physics majors (zero to nine), however, is comparable to that in the lecture courses for engineers, therefore the variability in students' number of peer interactions was likely sufficient. Final course grades, however, were less variable. While final grades ranged from C+ to A+ in the spring lecture course for physics majors, final grades only ranged from B to A+ in the fall offering (compared to a range of D- to A+ in the lecture course for engineers). The limited variability in student grades, therefore, are a plausible explanation for our results in the fall, but not the spring, offering of lecture course for physics majors.

Alternatively, these results may be due to a truly non-meaningful relationship between student interactions and final grade in this course. It is plausible that physics majors engage in peer interactions about their physics course because they are interested in the subject, and that the extent to which they engage in such interactions is more related to interest than performance. Future work, therefore, should aim to further understand the relationship between students' interaction network degree and their final grade in physics lecture courses across different class sizes, instructional styles, and student populations.

Expanding on these findings for our second and third research questions, we observed that the frequencies of the six different interaction topics were more uniform in the lecture courses than in the lab course. The most common interaction topic was \textit{homework}, followed by (in descending order) \textit{concepts}, \textit{lecture}, \textit{small-group work}, \textit{assessments}, and \textit{other} topics. Correspondingly, our statistical analysis indicated that \textit{concepts}, \textit{small-group work}, and \textit{homework} were the three interaction topics most strongly correlated with student performance, respectively. The remaining topics did not exhibit a significant relationship with final course grade. These results are consistent with Bruun and colleagues' demonstration of the significant impacts of peer interactions about both concepts and problem solving on performance, though we now explicitly link this pattern to different aspects of the course (i.e., problem solving is specifically about  student participation in small-group work during discussion sections)~\cite{bruun2013talking}.

Similar to the lab course, these findings for the lecture courses are likely attributable to a combination of the time allotted to each course component, the grading schemes, and the nature of the learning activities. Teaching concepts, for example, was the primary aim of the lecture courses and the central focus of all of the course components (small-group work, assessments, lecture, and homework). Students' interactions with peers about such concepts, in turn, were strongly related to their performance. Students also engaged in small-group work during two 50 min discussion sections each week and this participation accounted for 20\% of their final course grade in the lecture course for engineers. This small-group work was also collaboration-oriented, such that students were prompted to work together with their peers on physics problems. Interestingly, homework was an important interaction topic but only comprised 5\% of the final course grade in the lecture course for engineers (30\% in the lecture course for physics majors). We suspect the prevalence of interactions about homework is related to both time and the nature of homework assignments because students spent a handful of hours per week outside of class on the homework assignments, including attending homework help sessions where students often worked together. The homework problems were also similar to those on the exams (worth 65\% or 70\% of the final grade), providing student motivation to complete and understand them.

Surprisingly, though talking about lecture was not rare, such interactions were not strongly correlated with student performance. Lectures were the primary component of in-class time in both of the lecture courses (three 50 m sessions per week) and often implemented clicker questions discussed in small groups. Participation in lecture, however, hardly contributed to students' grades (5\% or 0\%). Still, this result is contrary to prior work showing that student engagement in in-class activities is linked to their conceptual understanding~\cite{smith2009peer}, which we would expect to be reflected in higher course grades, warranting future research. 

Finally, assessments were the least frequent interaction topic in the lecture courses and did not strongly correlate with students' grades. As mentioned above, this result could be due to students' interactions about concepts, rather than about the exams themselves, relating to their performance on exams, which comprise the majority of the final course grades (65\% or 70\%). Alternatively, this finding may be attributable to the timing of our data collection. The survey was not administered close to a midterm exam in all but one of the courses, which may explain the low frequency of this topic and weak correlation with overall performance. Future research should examine how the timing of such a survey impacts both the topics about which students report interacting and the relationship between interaction topics and final course grades.






\subsection{Limitations and future work}

We conclude by acknowledging the limitations to our study that motivate additional follow-up. First, our network survey may not have captured all interactions between students, for example due to recall bias where students do not remember a peer interaction and/or do not remember their peers' names to report. We also asked students about peers with whom they interacted ``this week." While this prompt likely captured many interactions that happen consistently week-to-week, it may also have captured one-off interactions that only occurred during the week of the survey. Future work should seek to disentangle these two kinds of peer interactions in terms of the number of peers students report and whether one kind of relationship is more impactful for student performance than another. Future studies should also explore the impact of different strategies for measuring students' interactions, such as providing full rosters of student names, and how these methods may affect the reported interactions and relationships with outcomes.

In our linear mixed models, we chose to simplify our data by only considering whether or not students interacted about each topic. This decision allowed our analysis to isolate the relationship between interaction topics and grades, intentionally removing information about the number of peers with whom students interacted about each topic. Future work should examine whether and how the number \textit{and} topics of peer interactions are related to student outcomes. For example, there could be thresholds of numbers of interactions above which the relationship with course grade plateaus and these thresholds may be different for each interaction topic.

In addition, as with any correlational analysis about student interactions, we cannot quantitatively disentangle whether peer interactions lead to higher grades or students with higher grades tend to interact more and/or about different topics. Future work to disentangle causation could probe student interactions over time, control for incoming course performance, or conduct student interviews about the roles of their various peer interactions.

We also examined the correlations between students' interaction topics and their course grades for all students combined. Prior work, however, has found mixed results as to whether students from different demographic groups engage in peer interactions to a similar extent~\cite{dokuka2020academic,dou2016beyond,williams2015understanding,wells2019,brewe2012investigating} and that course grading schemes can differentially impact the final grades of students from different demographic groups~\cite{webb2023attributing,paul2022percent,simmons2020grades}. Future work, therefore, should determine whether the role of interaction topics in student performance varies by student gender and race or ethnicity. Doing so would likely require much larger sample sizes than those included in this study, in order to have sufficient statistical power for multiple interaction terms. 

Finally, this study was conducted at one institution with only a few instructional styles. Future research should examine these relationships with different student populations and in different instructional contexts, particularly those with different course structures and grading schemes as these factors seem to be strongly related to interaction patterns.

\section{Conclusion}

We have built on the body of evidence suggesting that student interactions about different aspects of a physics course may be related to their performance in different ways. We also found that these relationships vary across instructional contexts: lab and lecture. Importantly, the interaction topics which most strongly correlated with students' grades mapped onto the assignments given the most weight in the course structures, through both the instructional time allotted and the grading schemes, and the nature of the assignments, whether collaborative or individually completed. These findings indicate that patterns of peer interactions, including which kinds of interactions are important for students' performance, are largely shaped by the instructional design of a course. Both instructors and researchers should consider these effects of course design on peer networks in order to ensure that all students have the opportunity to interact with their peers about meaningful topics that could impact their performance.

\section*{ACKNOWLEDGEMENTS}

This material is based upon work supported by the National Science Foundation Graduate Research Fellowship Program Grant No. DGE-2139899 (M.S. and M.D.) and Grant No. DUE-1836617. We thank Ashley Heim and Matt Thomas for engaging in meaningful discussions about this work.\\

\bibliography{InteractionsBibliography.bib}

\section*{Appendix}

\subsection*{ERGM coefficient estimates}

The coefficient estimates of the ERGM model for each network are summarized in Table~\ref{tab:coefficients}. We interpret the coefficient estimates as log-odds of edge formation. For example, the coefficient estimate for the \textit{homophily on lab section} variable for the fall offering of the lab course is 1.66. This means that the log-odds of an edge forming in the network increases by 1.66 for each additional edge connecting students in the same lab section, holding the rest of the network the same. In other words, edges connecting students in the same lab section are more probable than edges connecting students in different lab sections, even after accounting for the other configurations included in the model.

\begin{table}[ht]
\caption{\label{tab:coefficients}%
Coefficient estimates for our ERGM fit to the six observed networks. Standard errors of the coefficient estimates are in parentheses below. Asterisks indicate statistical significance ($^{*} p<$0.05; $^{**} p<$0.01; $^{***} p<$0.001).}
\setlength{\extrarowheight}{1pt}
\vspace{0.5cm}
\scriptsize
\begin{tabular}{lccccccc}
\hline
\textrm{}&
\multicolumn{2}{c}{Lab}&
\multicolumn{2}{c}{Lecture - Engineering}&
\multicolumn{2}{c}{Lecture - Physics}\\ 
\textrm{Variable}&
\textrm{Fall}&
\textrm{Spring}&
\textrm{Fall}&
\textrm{Spring}&
\textrm{Fall}&
\textrm{Spring}\\
\hline
\textit{Edges} & --7.73$^{***}$ & --9.07$^{***}$ & --6.62$^{***}$ & --8.17$^{***}$ & --2.72 & --7.12$^{***}$ \\
 & (0.55) & (0.50) & (0.53) & (0.38) & (1.52) & (1.86) \\
 \\
\textit{Reciprocity} & 1.39$^{***}$ & 2.12$^{***}$ & 4.43$^{***}$ & 5.17$^{***}$ & 2.87$^{***}$ & 2.05$^{***}$ \\
 & (0.22) & (0.18) & (0.24) & (0.20) & (0.47) & (0.62) \\
 \\
\textit{GWOD (outdegree distribution)} & --1.74$^{***}$ &  --1.44$^{***}$ & --1.47$^{***}$ & --1.32$^{***}$ & --1.04$^{*}$ & --1.22$^{*}$ \\
 & (0.19) & (0.14) & (0.22) & (0.18) & (0.50) & (0.59) \\
 \\
\textit{Homophily on lab section} & 1.66$^{***}$ &  3.65$^{***}$ & 0.15 & 0.72$^{***}$ & -0.85 & 1.08$^{*}$ \\
 & (0.17) & (0.10) & (0.24) & (0.16) & (0.96) & (0.52) \\
 \\
\textit{Homophily on discussion section} & 0.694$^{***}$ &  0.51$^{***}$ & 1.56$^{***}$ & 1.46$^{***}$ & 0.61$^{*}$ & 0.96$^{*}$ \\
 & (0.14) & (0.10) & (0.12) & (0.11) & (0.24) & (0.41) \\
 \\
\textit{Homophily on lab group} & 4.12$^{***}$ &  2.35$^{***}$ & 0.99$^{**}$ & 1.38$^{***}$ & 2.80$^{*}$ & 2.17$^{*}$ \\
 & (0.19) & (0.10) & (0.33) & (0.24) & (1.09) & (0.90) \\
 \\
\textit{Homophily on gender} & 0.42$^{***}$ & 0.38$^{***}$ & 0.50$^{***}$ & 0.56$^{***}$ & 0.37 & 0.86$^{**}$ \\
 & (0.10) & (0.08) & (0.12) & (0.10) & (0.24) & (0.33) \\
 \\
\textit{Main effect of gender on degree (woman)} & --0.03 & 0.11$^{**}$ & 0.10 & 0.09 & 0.55$^{**}$ & 0.28 \\
 & (0.05) & (0.04) & (0.06) & (0.05) & (0.18) & (0.17)\\
 \\
 \textit{Homophily on race or ethnicity} & 0.17 &  0.03 & 0.13 & 0.14 & 0.16 & 0.03 \\
 & (0.12) & (0.06) & (0.06) & (0.10) & (0.31) & (0.32) \\
 \\
\textit{Main effect of race or ethnicity on degree (URM)} & --0.03 &  0.02 & 0.17 & 0.08 & --0.16 & --0.13 \\
 & (0.11) & (0.06) & (0.13) & (0.07) & (0.33) & (0.34) \\
\\
 \textit{Main effect of final course grade on degree} & 0.19$^{**}$ & 0.23$^{***}$ & 0.17$^{*}$ & 0.24$^{***}$ & --0.17 & 0.40 \\
 & (0.07) & (0.06) & (0.10) & (0.05) & (0.20) & (0.23) \\
 \hline
\end{tabular}
\end{table}


\subsection*{ERGMs: Goodness of fit}

\begin{figure}
    \centering
    \includegraphics[width=3.6in,trim={0 0 9in 0}]{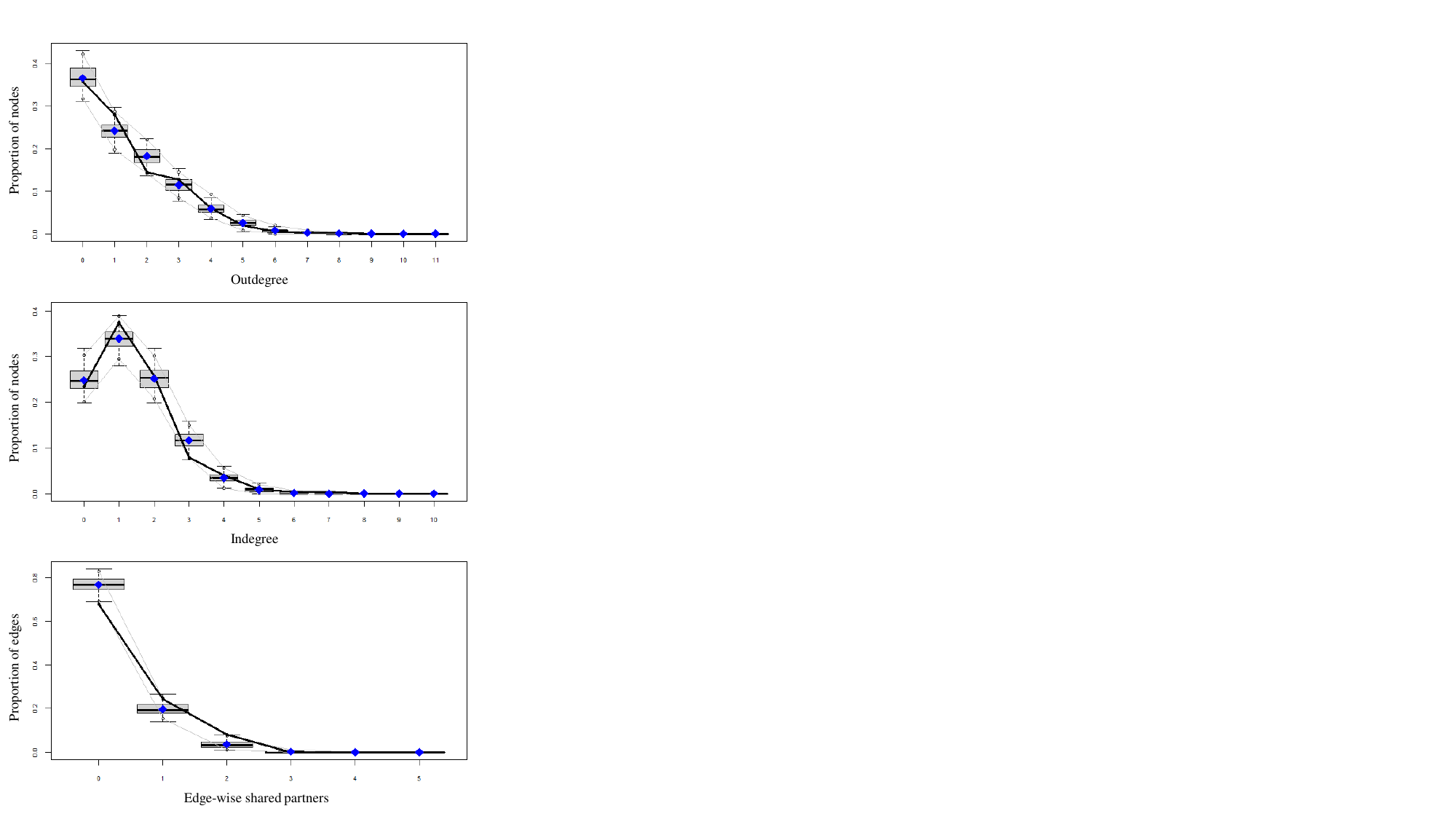}
    \caption{Goodness-of-fit plots for our ERGM applied to the peer interaction network of the fall offering of the lab course. The horizontal axes represent the value of each network measure: outdegree (top), indegree (middle), and edge-wise shared partners (bottom). The vertical axes represent the frequency of nodes or edges in the network with each value of the network measure. The black line represents our observed network and the box-and-whisker plots represent the distribution of 10 networks simulated using the ERGM coefficient estimates.}
    \label{fig:gofplots}
\end{figure}

The goodness-of-fit of an ERGM can be evaluated by comparing our observed network to a distribution of random networks simulated using the model coefficients. Figure~\ref{fig:gofplots} shows this comparison for one of the observed networks in this study for three different network measures: indegree (the number of incoming edges), outdegree (the number of outgoing edges), and edge-wise shared partners (measure of triadic closure, or small-group clustering). The boxplots represent the distribution of frequencies of these measures for 10 simulated networks. We see that our observed network, represented by the black line, falls within the distribution of simulated networks, indicating that our statistical model sufficiently represents the observed network. We observed similar goodness-of-fit plots for the remaining five networks in the study as well.

\subsection*{Linear mixed model results}

The coefficient estimates of both linear mixed models presented in the main text are provided in Table~\ref{tab:linearregresults}. Coefficients represent the mean increase in final course grade for a student interacting with at least one other peer about an interaction topic, as compared to a student who does not interact with any other peers about that topic. For example, a student who interacts with at least one other student about \textit{small-group work} in the lab courses have, on average, a final grade of 0.2 GPA points higher than a student who does not interact with any peers about \textit{small-group work}.

\subsection*{Model diagnostics for linear mixed models}

Here we assess the model diagnostics for the linear mixed models used in our analysis.

\subsubsection*{Variance inflation factors (VIFs)}

Variance inflation factors (VIFs) are a measure of multicollinearity of variables in a linear regression model, which can affect the model's precision. Multicollinearity indicates that two or more of the predictor variables vary closely with each other, reducing our ability to distinguish the significance of each variable on its own. VIFs measure the ratio of the standard error of a coefficient of a variable in the full model to the standard error of the coefficient of a variable in a model containing only that variable. For example, a VIF of two indicates that the standard error of a variable in the full model is twice what it would be in a model containing only that variable. VIF values less than two suggest adequate model precision and reliability. The VIFs of the predictor variables in our linear mixed models are all close to one (Table~\ref{tab:vifs}), suggesting sufficient precision of our estimated effects.  

\begin{table}[t]
\caption{\label{tab:linearregresults} Coefficient estimates for the linear mixed models. Standard errors of the coefficient estimates are in parentheses. Asterisks indicate statistical significance ($^{*} p<$0.05; $^{**} p<$0.01; $^{***} p<$0.001).}
\setlength{\extrarowheight}{2pt}
\vspace{0.5cm}
\centering
\begin{tabular}{lcc}
\hline
Predictor Variable & Lab & Lecture  \\ \hline
Concepts & 0.038 (0.030) & 0.189$^{***}$ (0.049) \\
Small-group work & 0.196$^{***}$ (0.030) & 0.160$^{**}$ (0.061) \\
Assessments & 0.043 (0.051)  & 0.037 (0.577) \\
Lecture & 0.089 (0.049)  & 0.089 (0.053) \\
Homework & 0.025 (0.034) & 0.150$^{**}$ (0.048) \\
Other & --0.017 (0.035) & --0.060 (0.061)\\
\hline
\end{tabular}
\end{table}

\begin{table}[t]
\centering
\caption{\label{tab:vifs} Variance inflation factors for the linear mixed models.}
\setlength{\extrarowheight}{1pt}
\vspace{0.5cm}
\begin{tabular}{lcc}
\hline
Predictor Variable & Lab & Lecture  \\ \hline
Concepts & 1.038 & 1.034 \\
Small-group work & 1.045 & 1.022 \\
Assessments & 1.029 & 1.011 \\
Lecture & 1.070 & 1.015 \\
Homework & 1.079 & 1.051 \\
Other & 1.009 & 1.017 \\
\hline
\end{tabular}
\end{table}

\subsubsection*{Checking model assumptions}

\begin{figure*}
    \centering
    \includegraphics[width=6.2in,trim={0 0 5.5in 0}]{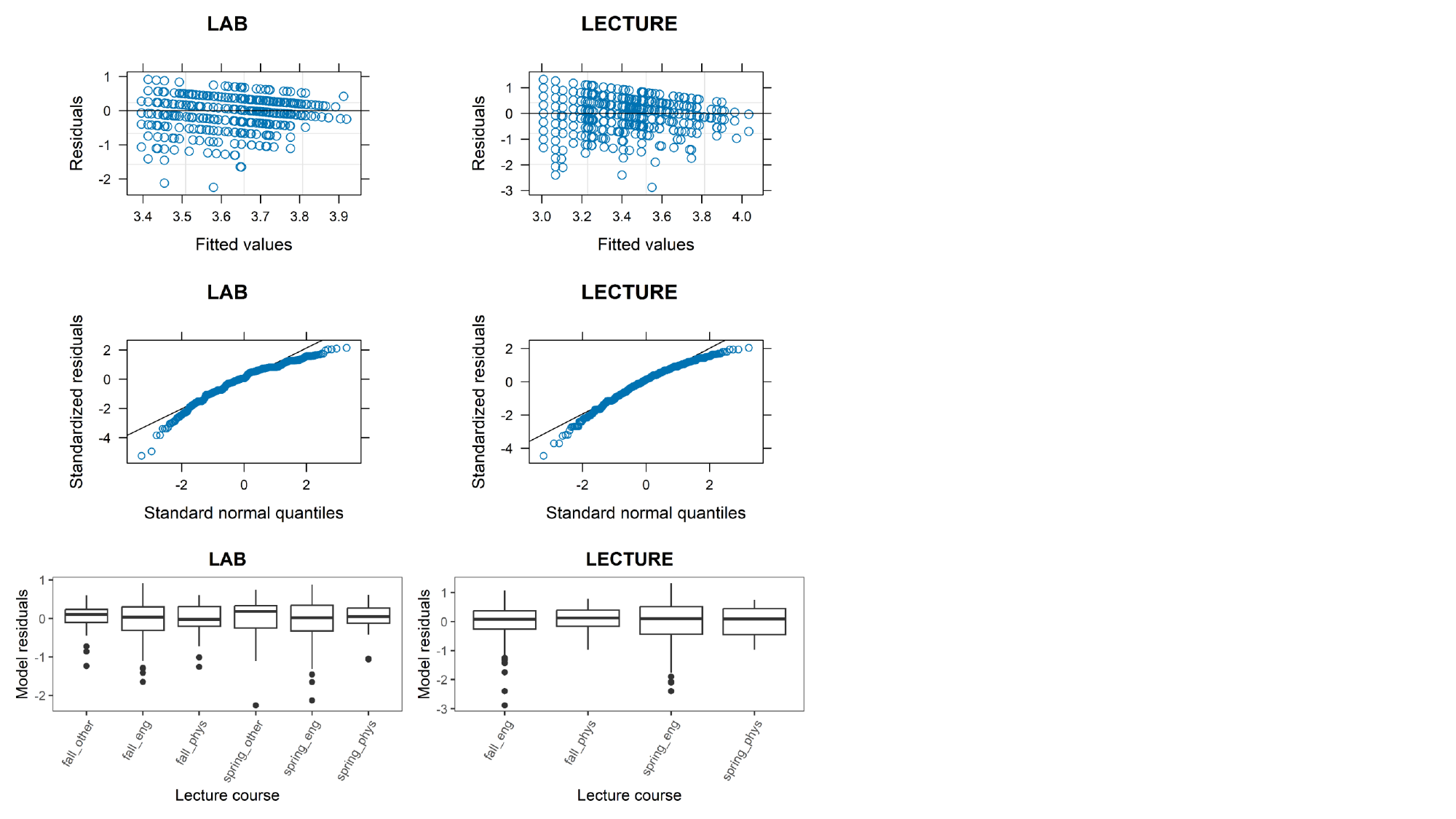}
    \caption{Model diagnostic plots for the linear mixed models. The plots in the top row visualize the residuals across the fitted values. The middle row shows quantile-quantile plots comparing the distribution of standardized residuals to a normal distribution. The bottom row shows the distribution of residuals at each value of the random effect variable (lecture course).}
    \label{fig:assumptions}
\end{figure*}

We checked the three main assumptions of linear mixed models: homoscedasticity of residuals, normality of residuals, and homogeneity of variance  of residuals (Fig.~\ref{fig:assumptions}). The assumption of homoscedasticity requires that residuals are randomly scattered about zero across the range of predicted values of the dependent variable. We do not observe any strong trends in the residual plots (top row of Fig.~\ref{fig:assumptions}), though we note that the discrete rows of residuals and the upper bound to the distribution are due to the nature of the dependent variable -- final course grades that are measured in discrete GPA values and capped at 4.3.  

Quantile-quantile plots are used to compare the distribution of residuals to a normal distribution. We mostly observe a normal distribution for our models, with some departure from normality at the tails of the distribution (middle row of Fig.~\ref{fig:assumptions}). This is a common pattern, however, and the regression results are still valid when the dependent variable is not normally distributed if the sample size is sufficiently large, as we have in this study~\cite{thomas2002normality}. 

Lastly, the homogeneity of variance assumption requires that there are not significant differences in the distribution of residuals for each value of the random effect variable, in this case the different lecture courses in which students were enrolled. The boxplots of the residuals within each lecture course show fairly consistent medians and interquartile ranges (bottom row of Fig.~\ref{fig:assumptions}). One-way ANOVAs comparing the residuals across the lecture courses also do not suggest a significant difference between the variances of residuals (lab: \textit{p} = 0.96, lecture: \textit{p} = 0.96).

\end{document}